\documentstyle[preprint,prl,aps]{revtex}
\draft
\newcommand{\be}{\begin{equation}} \newcommand{\ee}{\end{equation}}
\newcommand{\ben}{\begin{eqnarray}} \newcommand{\een}{\end{eqnarray}}

\begin{document}
\title{Early Universe Test of Nonextensive Statistics}
\author{Diego F. Torres\thanks{dtorres@venus.fisica.unlp.edu.ar}, 
H\'ector Vucetich, and 
A.\ Plastino\thanks{
plastino@venus.fisica.unlp.edu.ar}}
\address{Departamento de F\'{\i}sica, Universidad Nacional de La Plata\\
C.C. 67, C.P. 1900, La Plata, $ Argentina$}

\maketitle
\begin{abstract}
Within an early Universe scenario, 
nonextensive thermostatistics (NET) is investigated on the basis of
data concerning primordial Helium abundance.
We obtain first order corrections to the energy densities and weak 
interactions rates, and use them to compute the deviation in the
primordial Helium abundance. After compare with observational
results, a severe bound is stablished.

{\it PACS number(s):} 04.90. +e, 05.20. -y
\end{abstract}

\newpage

%\section{Introduction}

For a variety of physical reasons much work is nowadays devoted to
nonlinear
formalisms. Among them we can single out nonextensive
thermostatistics (NET)
\cite{t1}  as an active area of research. NET is based upon the
following two
postulates \cite{t2,t3}:

{\bf Postulate 1}.- {\it The entropy of a system that can be found
with
probability
$p_i$ in any of  $W$ different microstates $i$ is given by}

\be
S_q=(q-1)^{-1}\, \sum_{i=1}^W [p_i\,-\,p_i^q] \label{Sq},
\ee
{\it with  $q$ a real parameter}. We have a different statistics for
every
possible $q-$value. Of course,

\be
\sum_i p_i =1,
\ee
and it is easy to see that, for $q=1$, one regains the 
Boltzmann-Gibbs form \cite{t2}. The  resulting physics is
extensive
just for $q=1$. Otherwise we are led into the realm of
nonextensivity
\cite{t1,t2,t3,t4}.

{\bf Postulate 2}.- {\it An experimental measurement of an observable
$A$,
whose
expectation value in microstate $i$ is $a_i$, yields the $q$-
expectation
value} (generalized expectation value (GEV))
\be
<\,A\,>_q\,=\,\sum_{i=1}^W\,p_i^q \,a_i \label{gev},
\ee
{\it for the observable} $A$.

Unappealing as the above postulates may perhaps be considered, 
it should be strongly stressed that these two statements have the
rank of
{\it axioms}. As such, their validity is to be decided {\it
exclusively} not by vague discomfort feelings but by the comparison with
experiment of the conclusions to which they lead.
One such a test was recently reported in refs. \cite{P2V,Tsallis3}, where
bounds to $|q-1|$ were stablished using the
cosmic blackbody radiation. Also, present day determination
of Stephan-Boltzmann constant $\sigma$ puts a similar constraint,
for which the order of magnitud is $|q-1| <10^{-4}$.
However, it was later noted 
\cite{Barraco}, that in both cases,
the application of thermodynamics to these contexts is
strictly local, and thus, a non-violation of non-extensivity in a large
scale could not be sustained on this basis. In this communication, 
we intend to find bounds on non-extensivity not affected by such criticism,
which we deem fair.

The phenomenal success of thermodynamics and statistical physics
crucially
depends upon certain mathematical relationships involving energy and
entropy,
and much work has been devoted to i) show that many of these
relationships are
valid for {\it arbitrary} $q$, and ii) to find appropriate
generalizations for
the rest. In this vein we just mention that, by suitably maximizing
(\ref{Sq}), Curado and Tsallis \cite{t3} found that the whole
mathematical
(Legendre-transform based) structure of thermodynamics becomes
 {\it invariant} under a change of the $q-$value (from unity to any
other real number), while  the connection of NET both with quantum
mechanics
and with  Information Theory was established in \cite{t4}, where it
was shown
that all  of the conventional Jaynes-Boltzmann-Gibbs \cite{katz} 
results
generalize to  the Tsallis' environment.  For more details see
\cite{t1}.
Of course, to verify that NET is useful, it is necessary to show that
it
appropriately describes certain physical systems with $q$-values that
are
different from unity. Much work in this respect has been  been
performed recently. We may cite applications to
astrophysical problems \cite{t5,t6}, to L\'evi flights \cite{t7}, to
turbulence
phenomena \cite{t8}, to simulated annealing \cite{t9}, etc.
The interested reader is
referred to \cite{t1} for additional references.
Now, NET establishes a different (from the orthodox) fashion of doing 
`statistics', i.e., a non-conventional way of  {\it counting}, that has
proved to be useful in a variety of contexts. The
difference is governed by the value of the Tsallis parameter $q$.
It is clearly recommended to use  data concerning diverse natural
phenomena  to estimate the $q$-value with
reference to different scenarios. In the present communication we
shall attempt to use early Universe Helium abundance data so as to find
one of such estimates.

%\section{The notation}

For the canonical ensemble, (\ref{Sq}) gives \cite{t4}

\begin{equation}
\hat \rho=\frac{1}{Z_q}\left[ \hat 1 - (1-q) \beta \hat H\right]^\frac{1}{1-q},
\end{equation}
as the appropriate density operator and 

\begin{equation}
Z_q=Tr \left[ \hat 1 - (1-q) \beta \hat H\right]^\frac{1}{1-q},
\end{equation}
as the associated generalized partition function
\cite{t4}. Here, as usual, 
$\beta = 1/kT$ and $\hat H$ is the hamiltonian of the system.

We shall focus our attention upon the $\beta (q-1)\rightarrow 0$ limit, 
in which a first order expansion
allows for analytical computations. The expression of  the
generalized mean value 
of an operator was computed in this limit by Tsallis, Sa Barreto and Loh 
\cite{Tsallis3}. When applied to particle number operators, the
concomitant result reads 

\begin{equation}
\label{<n>}
<\hat n>_q= <\hat n>_{BG}  Z_{BG}^{q-1}
\left[ 1+ (1-q) x \left[ \frac{<\hat n^2>_{BG}}{<\hat n>_{BG}} +
\frac x2 \left( <\hat n^2>_{BG} - \frac{<\hat n^3>_{BG}}{<\hat n>_{BG}} \right)
\right] \right],
\end{equation}
where $x$ stands for $\epsilon/kT$ ($\epsilon$ is the energy of a
single 
particle) and the symbol $BG$ means {\sl to be computed within Boltzmann-Gibbs'
statistical tenets}.

%\section{Energy densities}

With the standard values of $<\hat n^2>_{BG}$ and $<\hat n^3>_{BG}$, 
for fermions and bosons, the corrections to the energy density
in the early universe may be computed \cite{Blaq}. When the particles are higly
relativistic, $T \gg m$, and non-degenerate $T \gg \mu$, we get

\begin{equation}
\rho_{bosons} = \frac{g_b}{2 \pi^2} \int_0^\infty dE E^3 <\hat n>_{bosons,q},
\end{equation}
\begin{equation}
\rho_{fermions} = \frac{g_f}{2 \pi^2} \int_0^\infty dE E^3 <\hat n>_{fermions,q},
\end{equation}
$g_{b,f}$ stands for the degeneracy factor of each one of the species
involved. Using (\ref{<n>}), we finally obtain

\begin{equation}
\label{rho}
\rho_{total}=
\rho_{bosons} +\rho_{fermions} = \frac{\pi^2}{30} g T^4 + \frac{1}{2\pi^2}
\left(  40.02 \, g_b+ 34.70 \, g_f\right)\,T^4 (q-1) ,
\end{equation}
where $g=g_b+7/8 g_f$. At  high enough temperatures, the energy
density of the universe is essentially dominated by $e^-,e^+,\nu$ and
$\hat \nu$. 
Interactions among these particles keep all of them at nearly 
 the same temperature. 
Accordingly, we set  $g_b=2$ and $g_f=2+2+2 \times 3$ and reach thus the final
form of the Tsallis correction to the energy density, namely 

\begin{equation}
\label{rho2}
\rho_{total}=
\rho_{standard} +21.63 \, T^4 (q-1).
\end{equation}

%\section{The weak rate}

We turn  now our attention to  the details of the 
computation of those corrections due to the weak interaction
rate. This rate allows one to compute the neutron abundance as the 
universe evolves.
We shall denote by $\lambda_{pn}(T)$ 
the rate for the weak processes to convert
protons into neutrons and by $\lambda_{np}(T)$  the rate for the 
associated, reverse ones.
Following  the  standard computations \cite{Weimberg,Bernstein}, 
it is 
possible to see that, for high enough temperatures,  the weak interaction
rate is $\Lambda (T) \simeq
\lambda_{np}+\lambda_{pn} \simeq G_F^2 T^5$, with 
$\lambda_{np}$ being related with $\lambda_{pn}$ by the principle of
detailed balance \cite{Bernstein}: $\lambda_{np}= \exp \left(-Q/T \right)
\lambda_{pn}$, $Q=m_n-m_p=1.29MeV$ and $G_F$ the Fermi constant. 
We want to compute first order Tsallis corrections ($(q-1)$-order) to
$\Lambda (T)$. To do this, we need to analyse the individual interaction
rates, i.e., each one of the terms in the sum

\begin{equation}
\lambda_{np}=\lambda_{\nu+n \rightarrow p+e^+}+
\lambda_{e^+ +n \rightarrow p+ \hat \nu}+
\lambda_{n \rightarrow p+e^- + \hat \nu}
\end{equation}
given by \cite{Weimberg},

\begin{equation}
\label{r1}
\lambda_{\nu+n \rightarrow p+e^-}=A \int_0 ^\infty dp_{\nu} p_{\nu}^2
p_e E_e (1-<\hat n_e>)<\hat n_{\nu}>
\end{equation}
\begin{equation}
\label{r2}
\lambda_{e^+ +n \rightarrow p+\hat \nu}=A \int_0 ^\infty dp_{e} p_{e}^2
p_{\nu} E_{\nu} (1-<\hat n_{\nu}>)<\hat n_{e}>
\end{equation}
\begin{equation}
\label{r3}
\lambda_{n \rightarrow p+e^- + \hat \nu}=A \int_0 ^{p_{0}}dp_{e} p_{e}^2
p_{\nu} E_{\nu} (1-<\hat n_{\nu}>)(1-<\hat n_{e}>),
\end{equation}
where $A$ is a constant fixed by the experimental value of
$\lambda_{n \rightarrow p+e^- + \hat \nu}$. In the preceding equations, 
we have to consider, of course,  energy conservation
as a boundary condition which relates $E_{\nu}$ and $E_{e}$.
Due to the high temperature limit we are interested in, several approximations
are in order
\cite{Bernstein}. 

i) All temperatures involved in the present game will be
taken as equal,  
$T_e=T_{\gamma}=T_{\nu}=T$, which ensures that reverse reactions
have the same form as the direct ones. 

ii) We shall neglect  Pauli factors
$(1-<\hat n_{\nu}>)$ and $(1-<\hat n_{e}>)$, and also the electron
mass in (\ref{r1}) and (\ref{r2}). 

With these approximations
(\ref{r1}) and (\ref{r2}) become identical. Using $p_e=E_e=Q+E_{\nu}$
in (\ref{r1}), the standard result follows. 

%\section{Our main result: a non-extensive bound for Tsallis' parameter}

Passing now to the nonextensive
context,  we must consistently use the $<\hat n>_q$ distributions
functions. Performing  the previous integrations, we obtain the leading order
corrections terms in the fashion

\begin{equation}                 
\lambda_{\nu+n \rightarrow p+e^-}=
\lambda_{\nu+n \rightarrow p+e^-}^{standard}+
\left( 480 T^5 +2 \times 84 T^4 Q + 18 T^3 Q^2\right)\,(1-q) A 
\end{equation}
where,

\begin{equation}                 
\lambda_{\nu+n \rightarrow p+e^-}^{standard}=
\left(4! T^2 + 2 \times 3!T Q +2! Q^2\right) \, A T^3.
\end{equation}

In order to get some fresh insight into the problem, we shall consider
here  only
the first correction,  
proportional to $T^5$.   
A more detailed analysis of these weak rates and the neutron--proton
abundance ratio they yield is considered elsewhere \cite{TORRES_T2}.
As explained in \cite{Bernstein}, the high temperature regime makes
$\lambda_{np} \simeq 2\times \lambda_{\nu+n \rightarrow p+e^-}$
and $\Lambda \simeq 2\lambda_{np}$. As a consequence, 
 the change in the weak reaction rate adopts the form  

\begin{equation}
\frac{\delta \Lambda}{A} =  1920 \; T^5 \;(1-q).
\end{equation}

We have now all the ingredients of the nucleosynthesis recipe at our
disposal. Basically,  nucleosynthesis 
is the competition between the weak interaction
rate and the expansion rate, given by the Hubble constant via the Einstein
equations.  The $^4\!He$ production may be estimated --in the standard
Big Bang model-- as 

\begin{equation}
Y_{p}=\lambda \left( \frac{2x}{1+x}\right) _{t_{f}} 
\end{equation}
where $\lambda = \exp (-(t_{nuc}-t_{f})/\tau )$ 
stands for the fraction of
neutrons which decayed into protons between $t_{f}$ and $t_{nuc}$, 
with $t_{f}$ $(t_{nuc})$ the time of freeze out of the weak 
interactions
(nucleosynthesis),
$\tau$
the neutron mean lifetime, and $x=\exp (-Q/kT)$ the neutron to proton
equilibrium ratio  \cite{Kolb}.            
It is  straightforward to compute the deviation produced in $Y_p$
by a variation in $T_f$, and correspondingly, $t_f$. We get 

\begin{equation}
\delta Y_p=Y_p\left[ \left( 1-\frac{Y_p}{2\lambda }\right) \ln \left( 
\frac{2\lambda }{Y_p}-1\right) +\frac{-2 t_f}{\tau _n} \right]
\frac{\delta T_f}{T_f},
\end{equation}
where  a 
radiation era relationship between time and temperature of the form 
$(T \propto t^{-\frac{1}{2}})$ is assumed and 
the one puts $\delta T_{nuc}=0$, because
it is fixed by the binding energy of the deuteron. 
Similar studies concerning bounds on gravitational theories were analized 
in \cite{Casas2,TORRES_N,TORRES_W}.
Considering now,
$Y_p=Y_p^{obs}=0.23$ and $\delta Y_p=0.01$, which is the observational
error \cite{yp}, and standard values for the times and the mean life
of neutron --which in fact, is not modified at order $(q-1)$--, we
must ask for 

\begin{equation}
\label{b2}
0.01 > 0.3766 |\frac{\delta T_f}{T_f}|
\end{equation}
to be satisfied in order to get an estimate of primordial
$^4\!He$ production compatible with observational data. 

In order to 
 obtain a value for $\frac{\delta T_f}{T_f}$ in nonextensive statistics,
we equate

\begin{equation}
\Lambda \simeq \left( \frac{\dot a}{a} \right)= 
\sqrt{\frac{8\pi G}{3} \rho_{total}}.
\end{equation} 

Adding up the corrections due to the changes in
the energy density and in the weak interaction
rate, our  first order result up to $(q-1)$ reads 

\begin{equation}
|\frac{\delta T_f}{T_f}|= 1276.4 \, (q-1),
\end{equation}
which allows for a stringent bound, using (\ref{b2}), on the value of $q$

\begin{equation}
\label{bound}
|q-1| < 2.08 \times 10^{-5}.
\end{equation}

In obtaining this bound we reach the main goal of the present communication:
the early universe physics places a bound upon the Tsallis parameter
$q$. 
Since the measured value of $Y_p$ comes from a sample
which has been thoroughly mixed, at least during the life of
the galaxy, and our estimate of $q$ is based on a volume of the order of
the horizon at nucleosynthesis epoch, the present test avoides
the locality problem pointed out in ref. \cite{Barraco}.
Of course, for other kinds of processes, different
bounds may be obtained. Our result pertains to those taking place during
the early childhood of our Universe.       
Thus, non-extensivity is severely constrained upon all
epochs of cosmic evolution, with separate, independently tested, 
observational evidence.

The authors acknoledge partial support from CONICET and valuable 
conversations with D.E. Barraco.

\end{document}